\documentclass{aa}
\usepackage{graphicx}
\usepackage{natbib}
\begin{document}
\title{ Relativistic magnetic reconnection at
 X-type neutral points}
   \author{Y.~Kojima\inst{\ref{inst1}}
           \and J.~Oogi
           \and Y.~E.~Kato
          }
%   \author{Yasufumi Kojima,
%           \inst{1}
%           Junpei Oogi
%     \and
%           Yugo E. Kato
%          }
%
\institute{
   Department of Physics, Hiroshima University, Higashi-Hiroshima
   739-8526, Japan \\
              \email{kojima@theo.phys.sci.hiroshima-u.ac.jp}\label{inst1}
             }
%
%   \date{}
%%%%
%
\abstract
  {Relativistic effects in the oscillatory damping of 
   magnetic disturbances near two-dimensional X-points
   are investigated.}
   {By taking into account displacement current, 
    we study new features of extremely magnetized systems,
    in which the Alfv\'en velocity is almost the speed of light.}
   {The frequencies of the least-damped mode are calculated
   using linearized relativistic MHD equations
   for wide ranges of the Lundquist number $S$
   and the magnetization parameter $\sigma$.}
   {The oscillation and decay times depend logarithmically on 
    $S$ in the low resistive limit. 
    This logarithmic scaling is the same as that for
    nonrelativistic dynamics, but the coefficient becomes
    small as $\sim \sigma^{-1/2}$ with increasing $\sigma$.
    These timescales approach constant values 
    in the large resistive limit:
    the oscillation time becomes a few times the light crossing time,
    irrespective of $\sigma$, and the decay time is proportional 
    to $\sigma$ and therefore
    is longer for a highly magnetized system.}
   {}
%\end{abstract}

 \maketitle
%------%-----%-----%-----%-----%-----%-----%-----%-----%

%(1)%%%%%%%%%%%%%%%%%%%%%%%%%%%%%%%%%%%%%%%%%%%
\section{Introduction}
%%%%%%%%%%%%%%%%%%%%%%%%%%%%%%%%%%%%%%%%%%%%%%%
%1
  The importance of magnetic reconnection manifests itself in
various energetic astrophysical phenomena, including 
relativistic objects such as
pulsars, magnetars, active galactic nuclei and gamma ray bursts. 
The characteristic propagation velocity 
for magnetic disturbances, the Alfv\'en velocity, 
depends on the magnetization parameter $\sigma$:
$2 \sigma$ represents the ratio of the magnetic to the rest mass energy 
density of the plasma. 
When the magnetization parameter $\sigma \gg 1$, the Alfv\'en velocity 
is almost the speed of light; 
the velocity becomes nonrelativistic in the opposite limit, 
$\sigma \ll 1$.
In this paper, we consider some inherently relativistic features that 
may appear in the magnetic reconnection when $\sigma$ is large.

In a simple analysis of the Sweet-Parker type reconnection,
the structure of the reconnection layer depends on two large
dimensionless numbers: $\sigma$ and the Lundquist 
(or magnetic Reynolds) number $S$, an inverse of 
resistivity (\cite{2003ApJ...589..893L}).
For $\sigma \ll S$, the inflow velocity is
nonrelativistic, and the reconnection is very similar to the classical
Sweet-Parker model. However, for 
$\sigma \gg S \gg 1$, the inflow
velocity becomes relativistic. 
\cite{2005MNRAS.358..113L} incorporated the compressibility of
matter and found that the inflow velocity is always sub-Alfv\'enic 
and remains much less than the speed of light, 
contradicting \cite{2003ApJ...589..893L}. The
reconnection rate is still estimated by substituting $c$ for 
the Alfv\'en velocity
in the nonrelativistic formula,
even in the relativistic regime.

%2
Small differences may also originate from the assumption of a steady state.
Numerical simulations of
an anti-parallel magnetic configuration in two dimensions have been 
performed without assuming a steady state
using relativistic resistive MHD code
(\cite{2006ApJ...647L.123W}), 
a relativistic two-fluid model(\cite{2009ApJ...696.1385Z}),
and PIC simulations on kinetic scale 
(\cite{2005PhRvL..95i5001Z,2007ApJ...670..702Z,
2008ApJ...677..530Z}).
See also \cite{2007MNRAS.382..995K} 
for the numerical schemes of the resistive relativistic MHD. 
These approaches have clearly demonstrated the relativistic 
dynamics, but simulation in a wide range of parameters 
would be time-consuming.
Moreover, the resolution becomes poor for small resistivity.

%3
The dynamics at an X-type null point, where a current sheet forms and the
magnetic energy is dissipated, have been studied previously. 
In context of nonrelativistic dynamics, \cite{1991ApJ...371L..41C} 
considered the
behavior of MHD waves near the X-point in the cold plasma approximation using
linear perturbation theory. They showed the remarkable result that the
dissipation time behaves as $ \sim (\ln S)^{2}$.
This logarithmic dependence, in contrast
to the normal power behavior $  \sim S^{\alpha}$, indicates fast decay.
Subsequently, the problem was studied 
analytically(\cite{1992ApJ...399..159H}) and by considering 
the propagation of linearized waves(\cite{2004A&A...420.1129M}).
Some physical properties of a more realistic system have also been
included, such as non-linear waves with 
thermal pressure(\cite{1996ApJ...466..487M, 2009A&A...493..227M}), 
electron inertial effects(\cite{2004ApJ...609..423M}), and 
viscosity(\cite{2005A&A...433.1139C, 2008A&A...487.1155C}). 
See a recent review of this topic given by \cite{2010SSRv..tmp...62M}
and references therein.

%4
The main concern of this paper is to explore relativistic effects
on the dynamical reconnection at an X-point.
In particular we consider whether the reconnection is qualitatively 
modified for a highly magnetized system with $\sigma \gg 1$, such 
as a magnetar.
We adopt a very simple system in order to understand the
differences, if any. Our work is a relativistic extension 
of \cite{1991ApJ...371L..41C}.
That is, we will calculate complex normal frequencies, which determine the
oscillatory damping of the magnetic disturbances with small
amplitudes, neglecting thermal pressure, viscosity and so on. 
The problem may be solved as an initial value problem, but
the initial data inevitably contain electromagnetic waves
besides MHD waves, and subsequent evolution may be complex.
In section 2, we discuss our numerical methods and boundary conditions. 
Our results are shown in section 3. Section 4 contains our conclusions.

%(2)%%%%%%%%%%%%%%%%%%%%%%%%%%%%%%%%%%%%%%%%%%%
\section{Model}
%%%%%%%%%%%%%%%%%%%%%%%%%%%%%%%%%%%%%%%%%%%%%%%
%(2.1)%%%%%%%%%%%%%%%%%%%%%%%%%%%%%%%%%%%%%%%%%
 \subsection{Equations for relativistic dynamics}
%%%%%%%%%%%%%%%%%%%%%%%%%%%%%%%%%%%%%%%%%%%%%%%
%1
We consider a two-dimensional problem, assuming $\partial /\partial z =0$.
In our model, the magnetic field $\vec{B} $ is located on a plane and the
electric field is perpendicular to it, $\vec{E} = E \vec{e}_z $. The electric
current $j \vec{e}_z $ is also perpendicular to the plane, and the charge
density consistently vanishes, since $\vec{\nabla} \cdot \vec{E} =0$. These
electromagnetic fields can be expressed in terms of only the
$z$-component of a vector potential $\vec{A} = A \vec{e}_z $ as 
\begin{equation}
\vec{B} = {\vec \nabla A } \times \vec{e}_z ,
 ~~~~~ E = - \frac{1}{c}\frac{\partial A }{\partial t } .
 \label{EBdef.eqn}
\end{equation}
The flux function $A$ satisfies with a wave equation with 
a source term: 
\begin{equation}
\left( -\frac{1}{c^2} \frac{\partial ^2 }{\partial t ^2 } + \nabla ^2
\right) A = -\frac{4 \pi j}{c} ,
  \label{Maxwell3.eqn}
\end{equation}
where the displacement current is included in contrast to the usual
nonrelativistic treatment.

%2
The dynamics of the plasma flow is determined by the continuity equation 
\begin{equation}
\frac{\partial \rho}{\partial t} + \vec{\nabla } \cdot ( \rho \vec{v} ) =0,
  \label{cont.eqn}
\end{equation}
and the momentum equation with the Lorentz force 
\begin{equation}
\rho \left( \frac{\partial}{ \partial t} + \vec{v} \cdot \vec{\nabla }
\right) \gamma \vec{v} = \frac{j}{c} {\vec e}_z \times {\vec B} = \frac{j}{c}
{\vec \nabla } A , 
  \label{eqn.motion}
\end{equation}
where $\gamma = ( 1- (v/c)^2 )^{-1/2}$,
$\rho$ is the mass number density in the laboratory frame, and 
the proper one is $\rho/\gamma$.
In eq. (\ref{eqn.motion}), the Coulomb force vanishes and thermal
effects in the pressure and internal energy are neglected
in the cold limit. 
This cold plasma approximation simplifies the problem:The slow 
magnetoacoustic wave is absent.
In non-relativistic dynamics, it is found that propagation of the 
fast one causes the current density to accumulate at the X point, 
where the energy is dissipated
\citep{2004A&A...420.1129M,2009A&A...493..227M}.
Thermal pressure is neglected, since our concern is the propagation 
in linearized system.
The finite pressure is meaningful in fully non-linear dynamics, 
where coupling and mode conversion between MHD waves are important 
in the neighborhood of the dissipation zone.

%3
Ohm's law with resistivity $\eta$ can be written as 
\begin{equation}
E + \frac{1}{c} (\vec{v} \times {\vec B} )_z = 
\frac{4 \pi \eta}{\gamma c^2} j ,
  \label{eqn.j1}
\end{equation}
which, in terms of $A$, is 
\begin{equation}
\left( \frac{\partial}{ \partial t} + \vec{v} \cdot \vec{\nabla } \right) A
= - \frac{4 \pi \eta}{\gamma c} j .
  \label{eqn.j2}
\end{equation}
The relativistic motion reduces the resistivity by the Lorentz factor 
$\gamma $. (See, e.g, \cite{1993PhRvL..71.3481B,2003ApJ...589..893L}.) 
However, this factor may be
set to $\gamma=1$ for a linear perturbation from a static background.

%(2.2)%%%%%%%%%%%%%%%%%%%%%%%%%%%%%%%%%%%%%%%%%
\subsection{Normal mode for the linearized system}
%%%%%%%%%%%%%%%%%%%%%%%%%%%%%%%%%%%%%%%%%%%%%%%
%1
We consider the dynamics of small perturbation 
in the vicinity of an X-point, which is governed by
current-free ($j_{0}=0$), static ($\vec{v} _{0}=0) $ 
background fields 
with uniform density ($\rho = \rho_{0}$).

The magnetic potential $A_{0}$ of the background
field can be written in the Cartesian $(x,y)$ or 
polar coordinates $(r, \theta)$ as 
\begin{equation}
A_{0} = \frac{B_0}{2L} ( -x^2+y ^2) = -\frac{B_0}{2L} r^2 \cos(2 \theta) ,
\end{equation}
where $L$ is a normalization constant for the length and $B_0$ is a constant
representing the magnetic field at $r=L$.

%2
The linear perturbation approximation for 
eqs.~(\ref{Maxwell3.eqn})-(\ref{eqn.j2}) 
reduces to a single equation for $\delta A$: 
\begin{equation}
\left( \eta \frac{\partial }{\partial t}
+\frac{(\nabla A_{0})^{2}}{4\pi \rho_{0}}\right) 
 \left( -\frac{1}{c^{2}}\frac{\partial ^{2}}{\partial t^{2}}
+\nabla ^{2}\right) \delta A
-\frac{\partial ^{2}}{\partial t^{2}}\delta A=0.
  \label{eqn.basic}
\end{equation}%
By using normalized length $\bar{r}=r/L$ and time $\bar{t}=v_{0}t/L$, 
where $v_{0}=B_{0}/(4\pi \rho _{0})^{1/2}$, 
eq.~(\ref{eqn.basic}) becomes 
\begin{equation}
\left( \frac{1}{s_{*}} \frac{\partial }
{\partial \bar{t}}+\bar{r}^{2}\right) 
\left( -\sigma \frac{\partial ^{2}}{\partial \bar{t}^{2}}
+\bar{\nabla}^{2}\right)
\delta A-\frac{\partial ^{2}}{\partial \bar{t}^{2}}\delta A=0,
  \label{eqn.basicN}
\end{equation}
where $s_{*}$ and $\sigma$ are non-dimensional parameters given by
\begin{equation}
 s_{*} =\frac{v_{0}L}{\eta },
\end{equation}
\begin{equation}
\sigma =\frac{B_{0}^{2}}{4\pi \rho _{0}c^{2}}
=\frac{v_{0}^{2}}{c^{2}}.
\end{equation}
The magnetization parameter $\sigma $ has been introduced 
through the displacement current, and hence eq.~(\ref{eqn.basic}) 
becomes eq.~(2.4) of \cite{1991ApJ...371L..41C} when the D'Alembertian 
$-\sigma \frac{\partial ^{2}}{\partial \bar{t}^{2}}+\bar{\nabla} ^{2}$ 
is replaced by the Laplacian $\bar{\nabla} ^{2}$ in the limit
of $\sigma =0$.
It should be noted that $v_{0}$
represents the Alfv\'en velocity at radius $L$ only in the 
nonrelativistic case.
The Alfv\'en velocity at $L$ is in general given by 
$V_{A}\equiv$ $ c \sigma ^{1/2} /(\sigma +1)^{1/2}$
$=v_{0}/(\sigma +1)^{1/2}$.
For highly magnetized cases where $\sigma \gg 1$, we have
$V_{A}\approx  c $, whereas $V_{A}\approx  v_{0} $ for $\sigma \ll 1$.
Although eq.~(\ref{eqn.basicN}) is used for mathematical calculation, 
the physical results are presented after normalization by $V_{A}$.
The Lundquist number $S$ characterizing 
the system is defined in terms of the Alfv\'en velocity 
$V_{A}$,  the radius $L$ and resistivity $\eta$ as 
\begin{equation}
S=\frac{V_{A}L}{\eta }.
\end{equation}
The related parameter $s_{*}$ is 
$ s_{*} = (\sigma +1)^{1/2} S$.
%%

%3
Equation (\ref{eqn.basicN}) exhibits two different 
behaviors near to and far from the origin. 
For large $\bar{r}$, the dissipating term with $s_{*}$ can be 
neglected, so that we have
\begin{equation}
\left[
  - \frac{ \sigma \bar{r}^{2} +1 }{\bar{r}^{2}}
    \frac{\partial ^{2}}{\partial \bar{t}^{2}}
   +\bar{\nabla}^{2}
\right] \delta A =0.
  \label{eqn.wave}
\end{equation}
This is exactly the equation in the cold plasma limit for the propagation of a 
fast magnetoacoustic wave,
whose velocity at $\bar{r}$ is given by the Alfv\'en velocity 
\begin{equation}  
 v_{A}(\bar{r}) \equiv 
\frac{v_{0}\bar{r}}{(\sigma  \bar{r}^2 +1 )^{1/2} }
= \frac{c \sigma^{1/2} \bar{r}}{(\sigma  \bar{r}^2 +1 )^{1/2} } .
\end{equation}
On the other hand, close to the origin, 
the term with $\bar{r}^2$ can be neglected in eq.~(\ref{eqn.basicN}).
After integrating by $\bar{t}$ once, we have
\begin{equation}
\left[
   -\sigma \frac{\partial ^{2}}{\partial \bar{t}^{2}}
   -\frac{1}{s_{*}} \frac{\partial } {\partial \bar{t}}
    +\bar{\nabla}^{2}
\right] \delta A =0.
  \label{eqn.tel}
\end{equation}
This is the so-called telegraphist's equation, in which
the effect of the finiteness of the velocity $c$ 
on the resistive losses, 
or the effect of resistivity on the wave equation, 
is taken into account. (See, e.g.,\cite{1953mtp..book.....M}.)
In the limit of $\sigma =0$, the equation becomes
the diffusion equation. 
Thus, eq.~(\ref{eqn.basicN}) leads to an
advection-dominated outer region described by eq.~(\ref{eqn.wave}) and a 
diffusion dominated inner one described by eq.~(\ref{eqn.tel}).
The diffusion region may be highly modified in nature for
large $\sigma$, as electromagnetic wave propagation becomes
important even in the diffusion zone for
a highly magnetized system. 
The critical radius $\bar{r}_{c}$, which separates the two regions, 
will be determined by the following normal mode analysis.

%4
  We solve eq.~(\ref{eqn.basicN}) as an eigenvalue problem in the form 
\begin{eqnarray}
\nonumber
\delta A &=& 
f(\bar{r}) \exp( im \theta ) \exp( -i {\bar{\omega}} \bar{t} )
\\
 &=& f(\bar{r}) \exp( im \theta )
    \exp( -i{\bar{\omega}} V_{A} (\sigma +1)^{1/2} t/L ),
  \label{eqn.Fform}
\end{eqnarray}
where ${\bar{\omega}} $ is a complex number. We only consider 
the axially symmetric $m=0$ mode, which is relevant to reconnection 
at the origin, as discussed in \cite{1991ApJ...371L..41C}. 
Another type of reconnection 
for $m \ne 0$ is discussed 
by \cite{1993ApJ...417..748O} and \cite{2005ApJ...632L.151V}, 
but that not is considered here.
Equation (\ref{eqn.basicN}) becomes 
\begin{equation}
\frac{1 }{ \bar{r} } \frac{d }{ d \bar{r} } \bar{r} \frac{d }{d \bar{r} } f
+{\bar{\omega}} ^2 
\left( \sigma + \frac{ 1} { \bar{r}^2 -i {\bar{\omega}} s_{*} ^{-1} } \right)
f =0 .  
  \label{eqn.Eigen}
\end{equation}
From this, a natural choice of the core radius $ \bar{r}_{c}$
is of order $( |{\bar{\omega}}|/s_{*} )^{1/2}$
$ \sim S^{-1/2}$ and $ \bar{r}_{c} $
corresponds to the usual skin depth  (\cite{1991ApJ...371L..41C}).
The dissipative term is dominant for 
$\bar{r} < \bar{r}_{c}$, whereas 
outside the critical radius eq.~(\ref{eqn.Eigen}) represents wave propagation, 
since the term with $|{\bar{\omega}} s_{*} ^{-1}| = \bar{r}_{c} ^2 $
can be neglected.
The current density is concentrated around the null point.

%5
A series solution inside the radius $\bar{r}_{c} $ may be expressed as 
\begin{equation}
f = 1 -\frac{1}{4} ( {\bar{\omega}}^2 \sigma 
+ is_{*} {\bar{\omega}}) \bar{r} ^2  + \cdots ,
  \label{eqn.center}
\end{equation}
where we have normalized to $f=1$ at the origin. We solve 
eq.~(\ref{eqn.Eigen}) with boundary condition (\ref{eqn.center}), 
from $\bar{r} =\bar{r} _c $ to $1$, assuming a complex number
${\bar{\omega}}$. The boundary condition imposed on the
circle $\bar{r} =1$ is $f =0$.
This means that the magnetic flux is frozen and
$\delta E =\delta j =\delta v =0$ there.
Thus, we have a one-dimensional eigenvalue problem 
for ${\bar{\omega}}$.

%6
Our main concern is not whole eigenfrequency spectrum, 
but rather the lowest frequency mode, which 
persists for a long time in the magnetic reconnection.
In particular, we will study the effect of the magnetization parameter 
on it. For this purpose, we first calculate $\bar{\omega}$ 
for the case $\sigma =0$, and then repeat the calculation,
gradually changing the parameter $S$ or $\sigma$.

%%(3)%%%%%%%%%%%%%%%%%%%%%%%%%%%%%%%%%%%%%%%%%%
\section{Results}
%%%%%%%%%%%%%%%%%%%%%%%%%%%%%%%%%%%%%%%%%%%%%%%
%1 
 The oscillation time $t_{osc}$ is defined 
in terms of the real part of the eigenfrequency ${\bar{\omega}} $ by 
$t_{osc}= 2\pi L/( (\sigma +1 )^{1/2} {\rm Re}({\bar{\omega}} ) V_{A})$.
(A factor $(\sigma +1)^{1/2}$ comes from our normalization of
${\bar{\omega}}$. (See eq.~(\ref{eqn.Fform}).)
Figure 1 shows the normalized time $V_{A}t_{osc}/ L$ 
as a function of $S$ for several values of $\sigma$.  
\cite{1991ApJ...371L..41C} showed that the relation 
$V_{A}t_{osc}/ L \approx 2 \ln S$ $ \approx 4.6 \log S$
holds for a wide range of $S$ with $\sigma =0$. 
The origin of this relation can be understood by considering the
traveling time of an MHD wave from the outer boundary to
the resistive region,
\begin{equation}
t_{osc} \sim  \int_{\bar{r}_{*}}  ^{1} 
 \frac{L }{v_{A}(\bar{r})} d\bar{r} .
  \label{eqn.travel}
\end{equation}
The velocity in the limit of $\sigma =0$ 
is scaled by $v_{A} \propto \bar{r}$, and
the dominant contribution in eq.~(\ref{eqn.travel}) comes from a 
small core region. By choosing the lower boundary $\bar{r}_{*}$
as $\bar{r}_{c}$, we have 
$t_{osc} \propto - \ln {\bar{r}_{c}} \propto \ln S $. 
%

%2
  When $\sigma $ is included, the oscillation time 
deviates from the relation 
$V_{A}t_{osc}/ L \approx 2 \ln S$.
The normalized time, in general, becomes
smaller than that at $\sigma =0$, as shown  in Fig.~1.
The logarithmic dependence with $S$ can be seen only
in the larger regime, and the coefficient 
in front of $ \ln S$ becomes smaller as $\sigma $ increases.
The Alfv\'en velocity becomes relativistic for $\sigma > 1 $
at the boundary, and approaches zero toward the center.
The velocity becomes nonrelativistic, at the radius
$ \bar{r} _{N} \approx  \sigma ^{-1/2}$ for $\sigma \gg 1 $, 
and the velocity is almost equal to $c$ outside this radius. 
The wave traveling time in eq.~(\ref{eqn.travel})
is almost determined by the slow region inside $\bar{r} _{N}$, and
the system size may be regarded as being effectively reduced 
to $\sigma ^{-1/2}L$.
We therefore have 
$V_{A}t_{osc}/(\sigma ^{-1/2}L)  \approx 2  \ln S$, 
i.e, $V_{A}t_{osc}/L  \approx 2  \sigma ^{-1/2} \ln S$
for the large $S$ regime.
This property can be seen 
from the curves around $\log S \approx 50 $ in Fig.~1,
except for $\sigma =10^4 $.
A factor of $(\sigma +1 )^{1/2}$ instead of $\sigma ^{1/2}$
may provide a better extension to $\sigma =0$, but 
a simple correction is used here.

%3
Figure 1 also shows that $V_{A}t_{osc}/L$ 
approaches a constant in the small $S$ regime,
for sufficiently large $\sigma $.
Asymptotically the value of this constant as $S \to 1$ 
is empirically 
$ V_{A} t_{osc} /L \approx c t_{osc} /L \approx 2.5$,
which is independent of $\sigma$, as far as $\sigma \ge 10^2 $. 
In our model, the core size increases as
${\bar{r}}_{c} \propto   S^{-1/2}$,
and hence the traveling time (\ref{eqn.travel})
becomes smaller with decreasing $S$, 
but the lower bound is
a few times the light crossing time for a region of size $L$.

%4
  The critical value $S_{c}$, which
discriminates between constant 
$V_{A}t_{osc}/ L $ for smaller $S$ and  
$V_{A}t_{osc}/ L  \propto \ln S $ for larger $S$, 
is given approximately by 
$ \ln S_{c} \sim \sigma^{1/2} $,
or $ \log S_{c} \sim 0.4 \sigma^{1/2} $.
The transition is not very sharp 
but the relation does give the approximate boundary
between two distinct behaviors.
Because $ \log S_{c} \sim 1 $ for
$\sigma =10$ and $ \log S_{c} \sim 40 $ 
for $\sigma =10^{4}$, which are located 
at the edges of Fig.~1, 
the two different behaviors are not clearly shown for these parameters.
This critical value $S_{c}$ also characterizes
a transition in the decay time as will be discussed below.

%%%%%%%%%%%%%%%%%%%%%%%%%%%%%%%%%%%%%%%%%%%%%%%
\begin{figure}[t]
%%%%%% FIG1 %%%%%
%
\centering
  \includegraphics[scale=0.5]{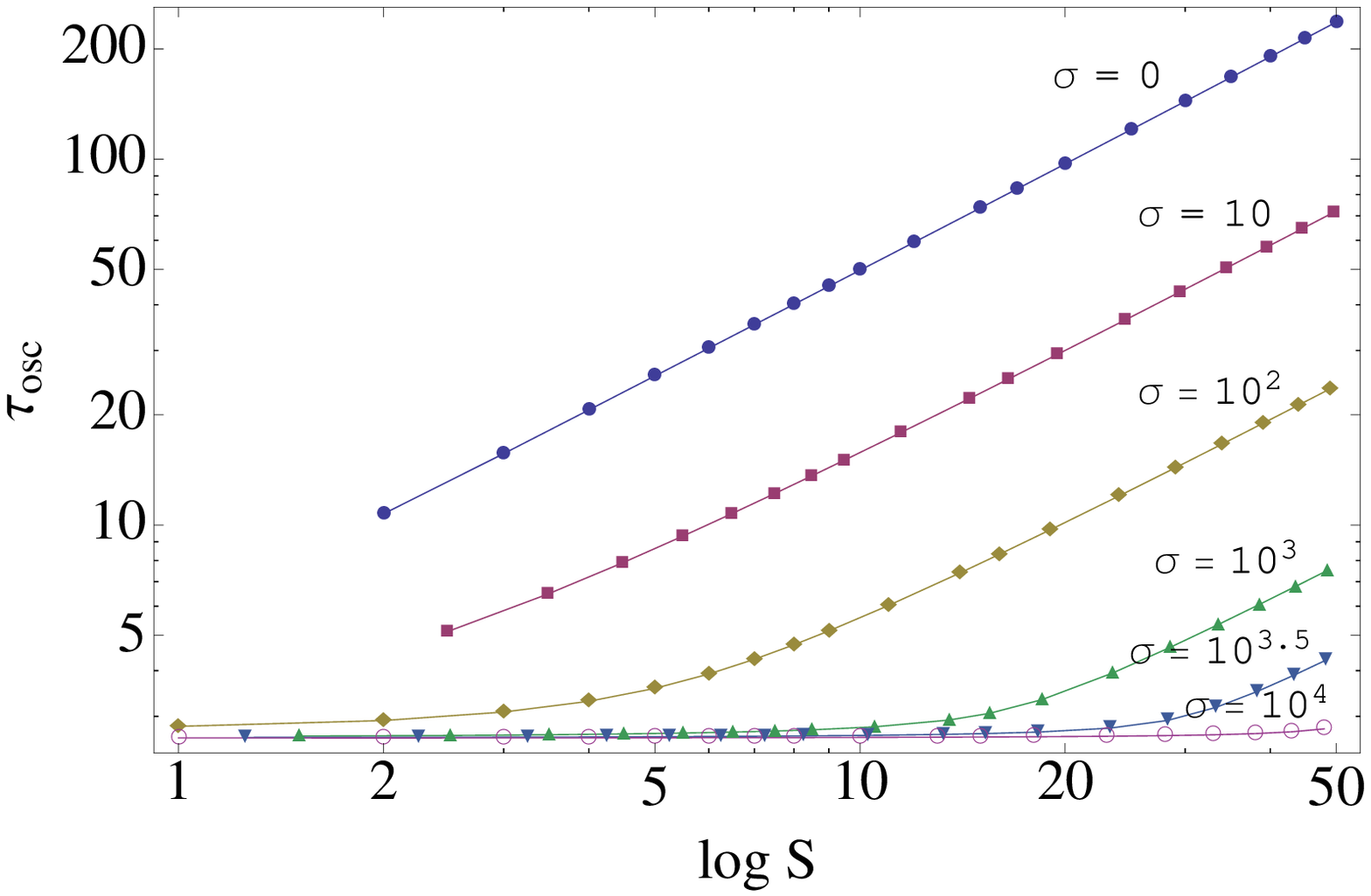}
\caption{ Normalized oscillation time 
$\tau _{osc} \equiv V_{A}t_{osc}/L $
as a function of Lundquist number $S$,
for magnetization parameter values
$\sigma = 0, 10, 10^{2}, 10^{3}, 10^{3.5} $ 
and $ 10^{4}$. 
}
%
%%%%% FIG2 %%%%%
%
\centering
  \includegraphics[scale=0.5]{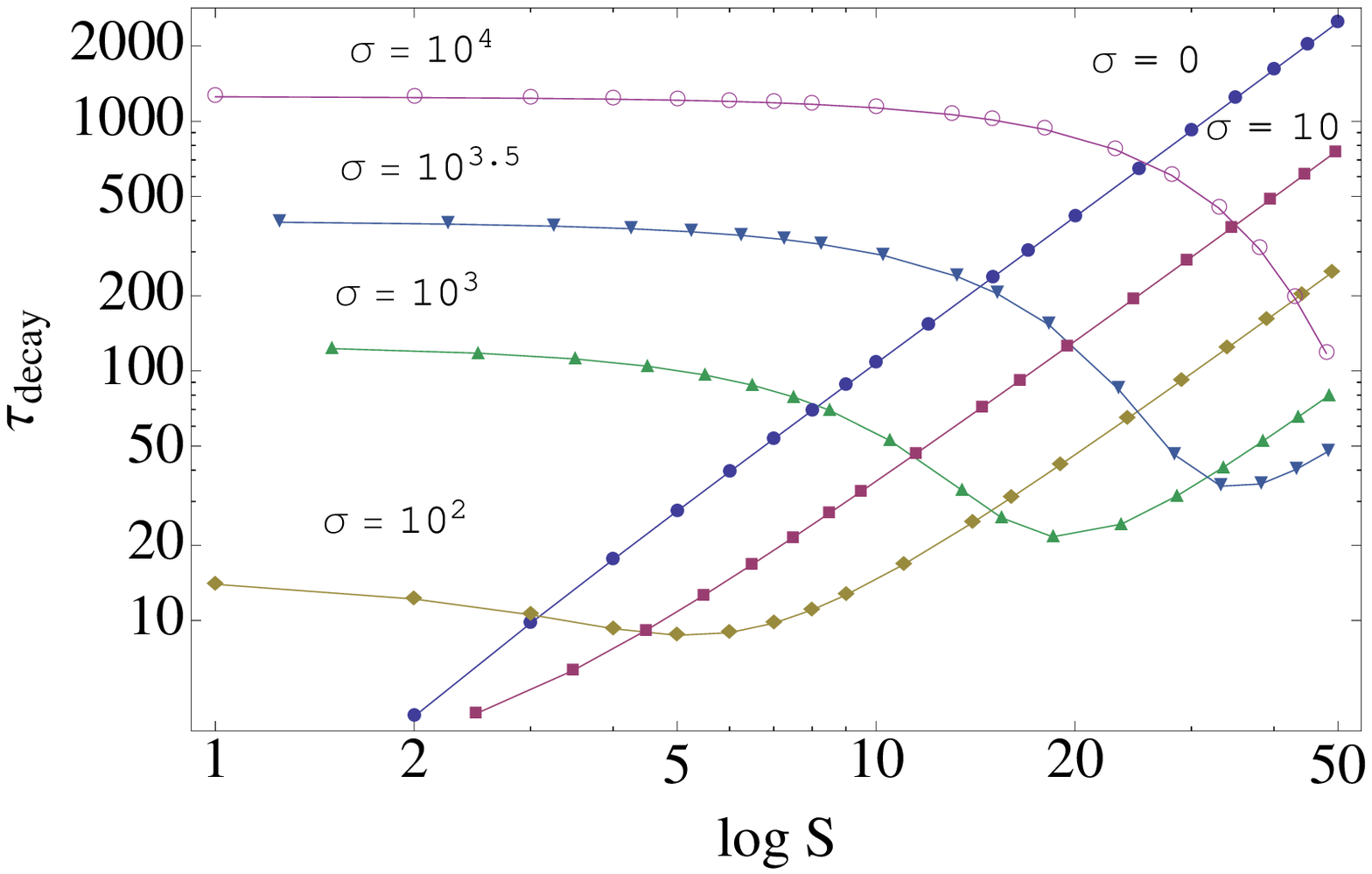}
\caption{ Normalized decay time 
$\tau _{decay} \equiv V_{A}t_{decay}/L $
as a function of Lundquist number $S$,
for magnetization parameter values
$\sigma = 0, 10, 10^{2}, 10^{3}, 10^{3.5} $ 
and $ 10^{4}$.
}
%%%%%%%%%%%%%%%%
\end{figure}
%%%%%%%%%%%%%%%%%%%%%%%%%%%%%%%%%%%%%%%%%%%%%%%

%%%
  The decay time is related to the imaginary part of
${\bar{\omega}}$,
$t_{decay}= L/( (\sigma +1)^{1/2} |{\rm Im}({\bar{\omega}})|V_{A})$.
Figure 2 shows the normalized decay time $V_{A}t_{decay}/ L$ 
as a function of $S$  for several values of $\sigma$.  
The time for $\sigma =0$ scales as
$ V_{A}t_{decay}/L $ $ = 2(\ln S)^{2}/\pi^{2}$ (\cite{1991ApJ...371L..41C}).
This scaling relation is also broken by the inclusion of $\sigma$.
The small and large $S$ regimes are different,
as they are for the oscillation time.
A typical example is given by the curve for $ \sigma =10^{3}$:
the critical value is $ \log S_{c} \sim 0.4 \sigma^{1/2} \sim 13 $ 
for this case.
Logarithmic dependence can be seen for $ \log S  > 20 $,
whereas the curve becomes constant for $ \log S  < 7 $.
The relation $ V_{A}t_{decay}/L \propto  (\ln S)^{2}$
can be seen in the large $S$ regime, $ S \gg  S_{c} $, 
except for $\sigma =10^{4}$, but 
the timescale is reduced to approximately 
$ V_{A}t_{decay}/L \approx  2\sigma ^{-1/2}(\ln S)^{2}/\pi^{2}$
for $\sigma \gg 1$.
The factor $\sigma ^{-1/2}$ can be interpreted as being due to an
effective reduction of the system's size, as considered 
for the oscillation time.
The normalized decay time becomes the minimum around  $S_{c}$.
%

%%%
 In the small $S$ regime,  $ S \ll  S_{c} $, 
normalized decay time approaches a constant value
$V_{A}t_{decay}/L \approx  0.14  \sigma $.
The normalized decay time for fixed $S$
increases with the magnetization parameter $\sigma $.
The limit of $\sigma \rightarrow \infty $ corresponds
to the vacuum, in which there is no matter ($\rho =0$)
and the dissipation time becomes infinite.  
This $\sigma $-dependence comes from taking account of the
finiteness of $c$ in the resistive losses. (See eq.~(\ref{eqn.tel}).)
This effect can be neglected 
in the large $S$ regime, where the
approximation of instantaneous dissipation is good.
However, the effect becomes evident in the small $S$ regime.

%------%-----%-----%-----%-----%-----%-----%-----%-----%
The energy $ E$ of perturbation  
decreases due to the Ohmic dissipation
\begin{equation}
\frac{d E}{dt}  = - \eta  \int j^2  dV. 
\end{equation}
The linearized form with Fourier component
provides an expression of the decay time as
\begin{equation}
\frac{ V_{A}  t_{\rm decay} }{L}
= 2 S \frac{
\int _{0} ^{1} (
 \delta  {\bar \varepsilon }_{B} + \delta  {\bar \varepsilon }_{E}
 + \delta  {\bar \varepsilon } _{M} ) 2\pi \bar{r} d\bar{r}
}{ 
\int  _{0} ^{1} |\delta  {\bar j} |^2  2\pi \bar{r} d\bar{r} } ,
\label{eqn.eng}
\end{equation}
where 
$ \delta  {\bar \varepsilon }$ is dimensionless energy density
of magnetic field, electric field, kinetic energy
of the fluid, and 
$ \delta  {\bar j} $ is dimensionless current density.
Their explicit forms are given by
\begin{equation}
 \delta  {\bar \varepsilon }_{B} =
 \frac{1}{8 \pi   }| \delta {\bar B} |^2  
= \frac{1}{2 } |  \frac{d f}{d {\bar r}  } |^2 ,
\label{def.EB}
\end{equation}
\begin{equation}
 \delta {\bar \varepsilon }_{E} = \frac{1}{8 \pi   }
|\delta {\bar E}|^2 
= \frac{\sigma }{2 } | \bar{\omega} f |^2 ,
\label{def.EE}
\end{equation}
\begin{equation}
  \delta  {\bar \varepsilon }_{M} 
= \frac{1}{2 }\rho _{0}|\delta {\bar v} |^2
 = \frac{\bar{r}^2}{2 | {\bar \omega} |^2 } 
| (\frac{1}{\bar{r}} \frac{d }{d {\bar r}  }\bar{r} 
\frac{d}{d\bar{r}} +{\bar \omega}^2  \sigma )f |^2 ,
\label{def.EM}
\end{equation}
and 
\begin{equation}
|\delta {\bar j} |^2  = 
| (\frac{1}{\bar{r}}
\frac{d }{d {\bar r}  }\bar{r} 
\frac{d}{d\bar{r}} +\bar{\omega}^2  \sigma )f|^2 .
\label{def.JJ}
\end{equation}
Spatial distributions of these energy densities
are displayed in Fig. 3 for $S=10^5$, $\sigma=10^1$ 
and in Fig. 4 for $S=10^5$, $\sigma=10^4$.
These functions are calculated by numerical solution
outside $ {\bar r}_{c}$, and by the analytic 
asymptotic form eq.~(\ref{eqn.center}) inside it.
Note that a sharp peak in $ \delta {\bar \varepsilon }_{M}$
and $ \delta {\bar \varepsilon }_{B}$
is located within $ {\bar r}_{c}$.
Both kinetic energy of matter and magnetic energy are 
accumulated from outer part to the core($\sim {\bar r}_{c}$),
and are dissipated in the central region.
However, distribution of electric energy is flat.
These overall features are not so much different
in Figs. 3 and 4, although the sharp peak shifts by
$ {\bar r}_{c} = ({\bar \omega}/((\sigma +1)^{1/2} S))^{1/2}$.

The magnitude of $ \delta {\bar \varepsilon }_{E}$
is much smaller than that of $ \delta {\bar \varepsilon }_{B}$
in Fig. 3 ($ \sigma =10^{1}$), 
whereas $ \delta {\bar \varepsilon }_{E}$
becomes comparable to $ \delta {\bar \varepsilon }_{B}$
in Fig. 4 ($ \sigma =10^{4}$). 
The electric energy is approximately proportional to
$\sigma$, as shown in eq.(\ref{def.EE}), and significantly 
contributes to the sum of energy.   
Hence, the decay time becomes longer with the increase 
of $\sigma$ for fixed $S$, since the total energy increases.
(See eq.(\ref{eqn.eng}).) 
In the large $S$ regime, however, 
the functions $ \delta {\bar \varepsilon }_{B}$
and $ \delta {\bar \varepsilon }_{M}$
are much larger than $ \delta {\bar \varepsilon }_{E}$,
so that the electric energy can be neglected.
The decay time does not increase with $\sigma$
in this  regime.
%------%-----%-----%-----%-----%-----%-----%-----%-----%

%%%%%%%%%%%%%%%%%%%%%%%%%%%%%%%%%%%%%%%%%%%%%%%%
\begin{figure}[h]
%%%%%% FIG3 %%%%%
%
\centering
  \includegraphics[scale=0.7,angle=270]{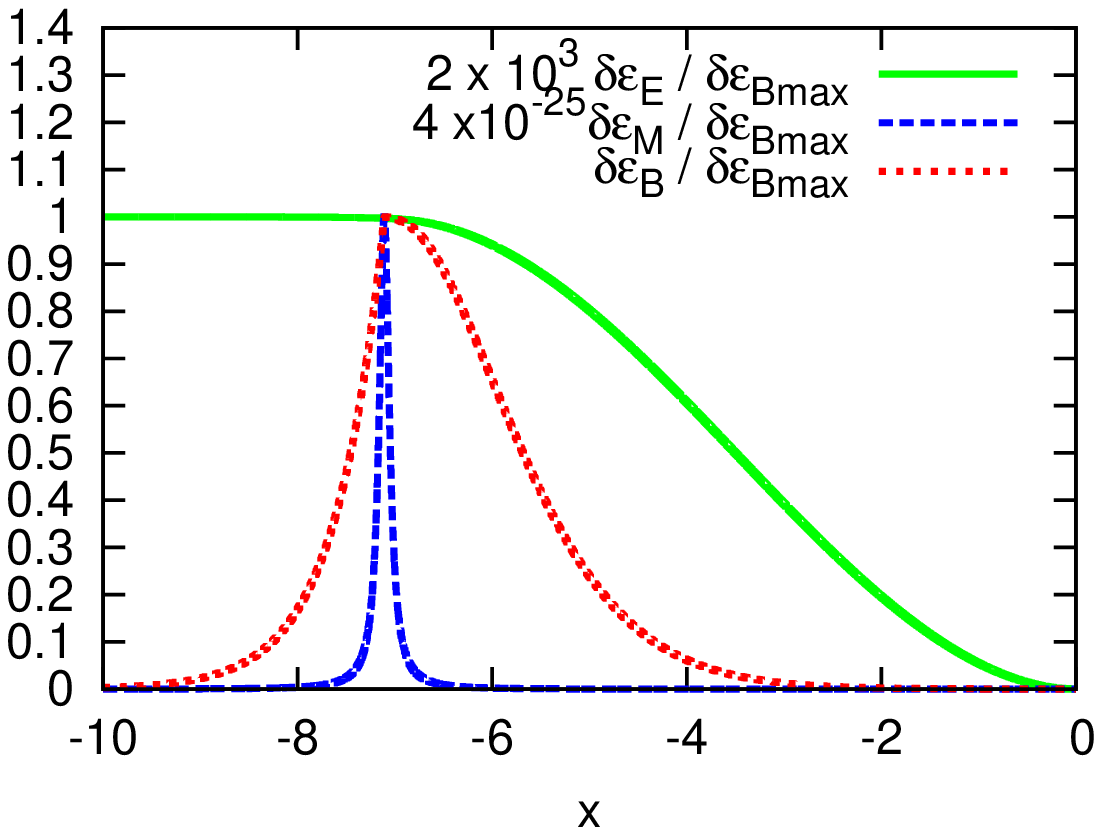}
\caption{ Normalized energy density
$\delta {\bar \varepsilon }$ as a function of
$ x= \ln {\bar r}$ for $ S=10^{5}$ and $\sigma = 10^{1}$.
The function $\delta {\bar \varepsilon }_{M}$
has a sharp peak, and is shown with a reduction factor
$4\times 10^{-25}$, 
while $\delta {\bar \varepsilon }_{E}$ is magnified
by $2\times 10^{3}$. 
}
%%%%% FIG4 %%%%%
\centering
  \includegraphics[scale=0.7,angle=270]{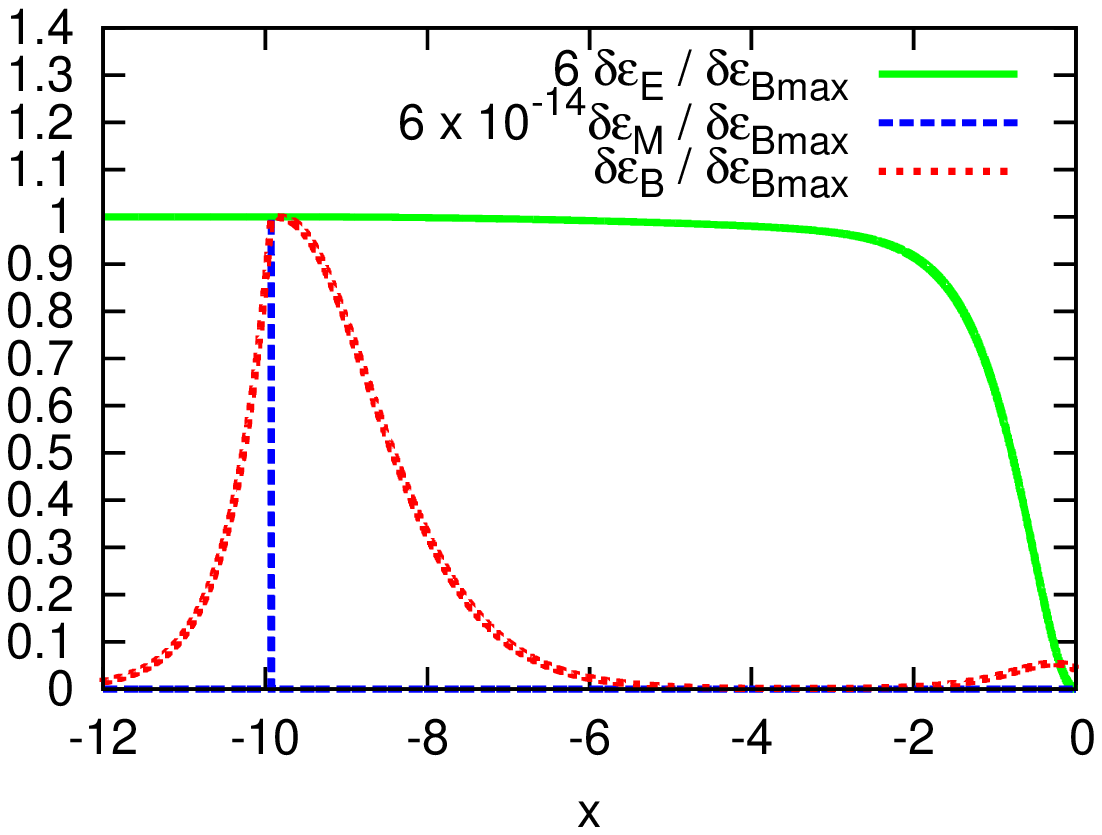}
\caption{ Normalized energy density
$\delta {\bar \varepsilon }$ as a function of
$ x= \ln {\bar r}$ for $ S=10^{5}$ and $\sigma = 10^{4}$.
The function $\delta {\bar \varepsilon }_{M}$
is shown with a factor $6\times 10^{-14}$, 
while $\delta {\bar \varepsilon }_{E}$ is shown with  
a factor $6$. 
}
\end{figure}
%%%%%%%%%%%%%%%%%%%%%%%%%%%%%%%%%%%%%%%%%%%%%%%%

%%(4)%%%%%%%%%%%%%%%%%%%%%%%%%%%%%%%%%%%%%%%%%%
\section{Discussion and conclusions}
%%%%%%%%%%%%%%%%%%%%%%%%%%%%%%%%%%%%%%%%%%%%%%%

%1
Relativistic MHD differs, in general, from the
nonrelativistic case in at least three ways: 
(i) the Lorentz factor $\gamma $, (ii) the Coulomb force $\rho _{e}E$, 
and (iii) the displacement current 
$ c^{-1} \partial E/\partial t$ in Maxwell's equation.
The Lorentz factor appears in the flow velocity and also in the resistivity of 
Ohm's law as a Lorentz contraction. The difference is of 
order $(v/c)^{2}$ in magnitude. Since we considered a linear
perturbation from the static state, the inflow velocity
is not very large and the Lorentz factor may approximate to 
$\gamma =1$.
The magnitude of $\rho _{e}E$ is of order $(v/c)^{2} $ 
times the Lorentz force $j \times B$, and is hence neglected in
nonrelativistic MHD. Moreover, the charge density is always zero 
due to the 2D X-point geometry considered here, so that 
the Coulomb force $\rho _{e}E$  vanishes exactly. 
This leaves the displacement current as a possible factor for the 
difference between relativistic and nonrelativistic MHD. 
We have studied its effects, especially on
the dynamics of the magnetic reconnection using
a simplified system based on
linearized equations in the cold plasma limit.
The magnetization parameter $\sigma $ is incorporated
in the basic equation through the displacement current and the
oscillation and decay times for the least-damped mode
were calculated numerically for 
parameters $S=$10-$10^{50}$ and $\sigma=$ 0-$10^{4}$.
%

%2%
In the system with $\sigma =0$, for which 
the displacement current can be neglected,
the oscillation and decay times are 
proportional to $\ln S$ and $(\ln S)^2$, respectively.
By including $\sigma $, these timescales are modified in 
different ways, in two regimes, 
which are characterized by 
$S \gg S_{c}$ or $S \ll S_{c}$ for $S_{c}\approx \exp(\sigma^{1/2})$.
For low resistivity, $S \gg S_{c}$,
a logarithmic dependence with $S$ can seen,
but the timescales normalized by the 
boundary radius $L$ and the Alfv\'en velocity $V_{A}$
become smaller with increasing $\sigma$.
The smaller timescales can be explained as being due to an
effective reduction in the size of the system, or the enlargement of 
the outer region 
where MHD waves propagate at almost the speed of light 
and the traveling time is negligible.
On the other hand, for high resistivity, $S \ll S_{c}$,
a new feature appears in both the oscillation and decay times,
which do not depend on $S$. 
The oscillation time is a few times the light crossing time
and does not depend on $\sigma$.
The dissipation time becomes longer in proportion to $\sigma $ 
and goes to infinity in the limit of $\sigma \rightarrow \infty $,
that is, no dissipation in the vacuum.
Reconnection at the X point is thought to be ``fast'', since the
dissipation time is scaled with $(\ln S)^{2}$.
Actual time is of the order of 10-$10^3$ times crossing time 
with Alfv\'en velocity.
The displacement current significantly spoils the good property, 
and the timescale increases with $\sigma$ in high resistive region.
The increase of the decay time is related with deficiency of matter, 
which is involved in the Ohmic dissipation.
Magnetic reconnection is expected to be an important process
of abrupt energy release in the solar and magnetar flares.
For example, the explosive tearing-mode reconnection
in the magnetar like the solar flares is discussed
(\cite{{2006MNRAS.367.1594L},{2010PASJ...62.1093M}}).
Dimensionless parameters are however quite different in them:
$\sigma \sim 10^{-4}$ and $S \sim 10^{14} $ in solar corona, 
whereas it is likely that $\sigma \gg 1$ and $S \gg 1 $ in a 
magnetar magnetosphere. 
Present result in an X-type collapse suggests the dissipation time 
$ t \sim 0.1 \sigma L/V_{A}$
$  \sim 10^{-5} \sigma $ ($L/10^{6}$cm) s
under highly magnetized environment.
The spiky rise time ($<0.1$s) or short duration ($<1$s) 
of the magnetar flare may significantly 
 constrain $ \sigma L$.
The energy of the flare $\Delta E $($\sim 10^{45}$ erg)
should be a part of magnetic energy within the volume $ L ^{3} $:
$B_{0} ^2 L ^{3} \sim \rho _{0} \sigma L ^{3} >\Delta E $.
These two conditions provide an upper limit of
$\sigma$ as
$ \sigma < 10^{4.5} (\rho_{0}/ {\rm (g /cm}^{3}))^{1/2}$. 
In such high energy events, radiation and possibly 
pair creation may  be important in the energy transfer.
Further study is needed for these effects.
However, the results in this paper demonstrate that the dynamics 
significantly depends on the  magnetization parameter through
the displacement current.
%%
%%%%%%%%%%%%%%%%%%%%%%%%%%%%%%%%%%%%%%%%%%%%%%%

%%%%%%%%%%%%%%%%%%%%%%%%%%%%%%%%%%%%%%%%%%%
\section*{Acknowledgements}
%%%%%%%%%%%%%%%%%%%%%%%%%%%%%%%%%%%%%%%%%%%%
This work was supported in part by a Grant-in-Aid for
Scientific Research (No.21540271) from
the Japanese Ministry of Education, Culture, Sports,
Science and Technology.

%%%%%%%%%%%%%%%%%%%%%%%%%%%%%%%%%%%%%%%%%
%-%  References 
%%%%%%%%%%%%%%%%%%%%%%%%%%%%%%%%%%%%%%%%%
%\bibliographystyle{aa}
%\bibliography{myrefs}
%--------------------------------------------

%%%%%%%%%%%%%%%%%%%%%%%%%%%%%%%%%%%%%%%%%
\end{document}